\documentclass[twocolumn]{autart} 
\usepackage{setspace}
\usepackage{amsmath}
\usepackage{amssymb}
\usepackage{bm}
\usepackage{bbm}
\usepackage{mathrsfs}
\usepackage{amscd}
\usepackage{dsfont}
\usepackage{graphicx}
\usepackage{epsfig}

\def\qed{\hfill \vrule height 7pt width 7pt depth 0pt\medskip}
\def\beq{\begin{equation}}
\def\eeq{\end{equation}}

\def\proof{\noindent{\bf Proof}\ \ }

\newtheorem{theorem}{Theorem}
\newtheorem{definition}[theorem]{Definition}
\newtheorem{proposition}[theorem]{Proposition}

\newtheorem{corollary}[theorem]{Corollary}
\newtheorem{example}{Example}

\newcommand{\ba}{\begin{array}}
\newcommand{\ea}{\end{array}}

\newcommand{\be}{\begin{equation}}
\newcommand{\ee}{\end{equation}}

\newcommand{\mc}{\mathcal}

\newcommand{\Z}{\mathbb{Z}}

\newcommand{\1}{\mathbbm{1}}
\newcommand{\E}{\mathbb{E}}

\newcommand{\R}{\mathbb{R}}
\newcommand{\N}{\mathbb{N}}

\renewcommand{\P}{\mathbb{P}}

\def\1{\mathds{1}}
\def\Z{\mathbb{Z}}
\def\N{\mathbb{N}}
\def\E{\mathbb{E}}
\def\R{\mathbb{R}}

\def\P{\mathbb{P}}

\usepackage{color}
\newcommand{\jcd}[1]{ #1}
\newcommand{\ff}[1]{#1}

\journal{}

\begin{document}

\begin{frontmatter}

\title{The robustness of democratic consensus}

\author[Torino]{Fabio Fagnani}
\author[UCL]{Jean-Charles Delvenne}
\address[Torino]{Department of Mathematical Sciences, Politecnico di
Torino, Corso Duca degli Abruzzi, 24, 10129 Torino Italy {\tt\small fabio.fagnani@polito.it}}
\address[UCL]{Corresponding author. Institute for Information and Communication Technologies, Electronics and Applied Mathematics (ICTEAM), Namur Complex Systems Center (naXys) and Center for Operations Research and Econometrics (CORE), Universit\'{e} catholique de Louvain, 4 Avenue Lema\^{i}tre, B-1348 Louvain-la-Neuve,
Belgium {\tt\small jean-charles.delvenne@uclouvain.be}, +32 10 47 80 53}%

\begin{abstract}
In linear models of consensus dynamics, the state of the various agents converges to a value which is a convex combination of the agents' initial states. We call it democratic if in the large scale limit (number of agents going to infinity) the vector of convex weights converges to 0 uniformly.
 
 Democracy is a relevant property which naturally shows up when we deal
 with opinion dynamic models and cooperative algorithms such as consensus over a network:
 it says that each agent's measure/opinion is going to play a negligeable
 role in the asymptotic behavior of the global system. It can be seen as a relaxation of average consensus, where all agents have exactly the same weight in the final value, which becomes negligible for a large number of agents.

 We prove that starting from consensus models described by time-reversible stochastic matrices, under some mild technical assumptions,
 democracy is preserved when we perturb the linear dynamics in finitely many vertices.
 We want to stress 
 that the local perturbation in general breaks the time-reversibility of
 the stochastic matrices. The main technical assumption needed in our result is the
 irreducibility of the large scale limit stochastic matrix\jcd{, i.e. strong connectedness of the limit network of agents,} and we show with an example that
 this assumption is indeed \jcd{required}. 
\end{abstract}


\end{frontmatter}

\section{Introduction}\label{intro}


\subsection{Consensus}
Many opinion dynamics models \cite{Golub}, \cite{Jackson} and cooperative algorithms over networks like consensus \cite{Jadba}, \cite{Olfati}, \cite{Carli}
are mathematically
represented by a stochastic matrix  $P\in\R^{V\times V}$ where $V$ is a finite set. Interpreting $x_i$ as an initial belief/opinion of agent $i\in V$ on
some fact or event, or a position in a physical space, linear consensus dynamics consists  in replacing each opinion  $x_i$ by a weighted average of the opinion of  agent $i$'s neighbors in the network. Such dynamics may be expressed by the equation $x(t+1)=P x(t)$ where $P$ is row-stochastic, i.e., is nonnegative with every row summing to one. 
\jcd{Another motivation is the design or analysis of agents such as robots moving on the real line or any Euclidean space, while exchanging messages on their respective positions in on a communication network. The equation $x(t+1)=P x(t)$ now describes the situation where  every agent  moves to a weighted  average of the position of their neighbors in the network. The robots typically seek to solve the consensus problem, i.e. to all reach a common position in the space.}


It is well known that under suitable assumptions on $P$
(\jcd{i.e.} irreducibility and aperiodicity) there exists  $\pi\in \R^V$ such that
\beq\label{Markov-convergence}\lim\limits_{t\to
+\infty}(P^tx)_i\to \sum_{j\in V}\pi_jx_j(0)\,,\quad\forall i\in
V\,.\eeq 
Moreover, $\pi_i>0$ for all $i \in V$, $\sum_i\pi_i=1$ and $\pi^*P=\pi^*$, \jcd{where $\pi^*$ denotes the transpose of $\pi$ and is thus a row vector}.

In terms of consensus or opinion dynamics, convergence (\ref{Markov-convergence}) means
that \jcd{the} opinion of all agents tends to the common value $\sum_{j\in
V}\pi_jx_j$ which is a convex combination of the initial opinions. For this reason, in this paper, 
a stochastic matrix $P$ for which (\ref{Markov-convergence}) holds will be called a {\it consensus matrix} and the relative vector $\pi$ the corresponding {\it consensus weight vector} of $P$.
If $\pi$ is the uniform vector (i.e. 
$\pi_i=|V|^{-1}$ for all $i$), the common asymptotic value is
simply the arithmetic mean of the initial beliefs; in other terms,
all agents equally contribute to the final common belief. This uniformity condition  amounts to assuming
that the matrix $P$ is doubly stochastic (also all columns sum to $1$), a sufficient condition for this being that $P$ is symmetric. 

In this paper we want to consider the situation where we have a
sequence $P^{(n)}$ of consensus matrices over a state space $V_n$ of
increasing cardinality corresponding to larger and larger sets of
interacting agents. The corresponding consensus weight vectors will be denoted by $\pi^{(n)}$.

\subsection{Democracy}

The sequence $P^{(n)}$ of consensus matrices is
called {\it democratic} if their corresponding invariant probabilities $\pi^{(n)}$ are such that $||\pi^{(n)}||_{\infty}:=\max_{i\in
V_n}\pi^{(n)}_i\to 0$ for $n\to +\infty$. This says that even if
the initial opinion of the various agents may have a different
weight on the final consensus value, still the weight of each of
them becomes negligeable as the number $n$ of agents grows to
$\infty$. This property has already been proposed in \cite{Golub}, \cite{Jackson} as `wise society' with the following interpretation. If
we assume that \ff{the initial opinion of the various agents are of type} $x_i=\mu+N_i$, where $\mu\in\R$ is the value of a
parameter we want to estimate and $N_i$ are independent noises
having mean $0$ and variance $\sigma_i^2$, then, \ff{the consensus point reached by applying the consensus matrix $P^{(n)}$ is given by}
$$\sum_j\pi^{(n)}_jx_j=\mu+N\,,\quad {\rm with}\;
N=\sum_j\pi^{(n)}_jN_j$$ If $\sigma_i^2$ are bounded from above,
it follows from a straightforward variation of the weak law of
large numbers \cite{Golub} that democracy implies that $N\to 0$ in
probability when $n\to +\infty$. In wise societies agents'
asymptotic belief converge to the real value of the parameter when
the number of agents goes to $\infty$.

A very special case is when we start from a sequence $G^{(n)}$ of connected undirected graphs (with no self loops) on the set of vertices $V_n$ and the consensus matrices $P^{(n)}$ are obtained by assigning homogeneous weights to all neighbors of an agent. \ff{Put $d^{(n)}_i$ equal to
the degree in $G^{(n)}$ of the vertex $i$ (number of edges
connected to $i$) and define
\begin{equation}\label{SRW}P^{(n)}_{ij}=\frac{1-\tau}{d^{(n)}_i}\quad\hbox{\rm  for $j$ neighbor of $i$}, \quad P^{(n)}_{ii}=\tau\end{equation}
while $P^{(n)}_{ij}=0$ if $j\neq i$ is not a neighbor of $i$ in  $G^{(n)}$,
where \jcd{$0 \leq \tau <1$} is a self-confidence parameter \jcd{(see e.g. \cite{Frasca,Friedkin} for other models of self-confidence, or stubborness, in opinion dynamics)}}. 
In this case we have that $\pi^{(n)}_i=d^{(n)}_i/\sum _jd^{(n)}_j$.  In this context,  democracy thus happens to be a rather easily checkable property only depending on the degrees of the various nodes.  In particular, if graphs are regular ($d^{(n)}_i$ constant in $i$)
the consensus weight vectors all coincide with the uniform one. More
generally, if we have a uniform bound $d^{(n)}_i\leq d$ for all $n$ and $i\in V_n$, then, clearly, $||\pi^{(n)}||_{\infty}$
goes to $0$.
This example is encompassed by the more general time-reversible consensus matrices which will be revised in next section. For them, 
an explicit characterization of the consensus weight vectors remains available so that
$||\pi^{(n)}||_{\infty}$ can be estimated and democracy can easily be checked. Quite a different story is when time-reversibility is lost (e.g. sequences $P^{(n)}$ constructed as in (\ref{SRW}) over directed graphs $G^{(n)}$): in this case there is no general techniques available to characterize the consensus weights vectors and check democracy.

In \cite{Golub} the authors propose a sufficient condition for democracy (see their Theorem 1) which can be applied also to stochastic matrices which are not time-reversible. However, one of their assumptions (Property 2) never holds when the underlying sequence of graphs have a bounded degree and this rules out many interesting examples.


\subsection{Robust democracy and main result}

In this paper we focus on the robustness of democracy with respect to local perturbations.
More precisely, we start from a democratic sequence 
$P^{(n)}$ defined on a sequence of nested sets $V_n$ of nodes (i.e., $V_n \subset V_{n+1}$) and we analyze what happens to the consensus weights vectors when
$P^{(n)}$ is locally perturbed. The perturbed 
sequence of consensus matrices $\tilde P^{(n)}$ coincides with $P^{(n)}$ but in a fixed finite
number of rows corresponding to a subset of vertices $W$.

Our Theorem \ref{theoweakdemocracy} shows that under very
mild assumptions (irreducibility of the limit chains\jcd{, i.e. strong connectedness of the limit graph}) $\tilde
P^{(n)}$ maintains a weak form of democracy (pointwise convergence
to $0$ of the consensus weight vectors). Afterwards, we focus on
time-reversible chains $P^{(n)}$ and in Theorem \ref{theorempertdemtr} we prove that, under some
technical assumptions (essentially that degrees are bounded in the associated graphs) the perturbed sequence $\tilde P^{(n)}$ (possibly no longer time-reversible) remains democratic. \ff{We again want to stress the fact that the sufficient conditions for democracy proposed in \cite{Golub} can not be applied in this context as their property 2 will never be satisfied.}
The proofs of these results will be probabilistic in nature interpreting $P^{(n)}$ and $\tilde P^{(n)}$ as transition matrices of Markov chains and the corresponding consensus weights as invariant probability vectors. Although our motivation and applications for our results lie in the field of opinion dynamics and consensus, we find the dual language of Markov chains more convenient and powerful to express the technical results and proofs. 

\subsection{Applications and context}

From the point of view of opinion dynamics, these results
essentially say that in democratic chains, no single agent or a
finite group of them can unilaterally break democracy by modifying
their outgoing links or weights as long as the number of links
remains bounded and the graph connected. 

As a more specific example, we can consider a sequence of connected undirected graphs over a nested set of vertices $V_n$ and $P^{(n)}$ to be the corresponding consensus matrices as defined in (\ref{SRW}). Fix now a subset $W\subseteq V_1$ and perturb $P^{(n)}$ on $W$ by assuming that agents in $W$ form a small community which is incline\jcd{d} to give more credit to each other'\jcd{s} opinion than to people outside of $W$.
This can be modeled by simply assuming that, for each $i\in W$,  all weights $\tilde P^{(n)}_{ij}$ for $j\in W$ are a factor $\lambda \jcd{\geq} 1$ greater than weights $\tilde P^{(n)}_{ij}$ for $j\jcd{\notin} W$.
The parameter $\lambda$, called `homophily', measures the `closure' of the community $W$ to external influence. Our results assert that, disregarding how large $\lambda$ is, democracy is preserved: in the final consensus the opinion of these agents still plays a negligeable role when $|V_n|\to +\infty$. This example is treated in a more formal way in Section 
\ref{sec-formulation} (see Example \ref{main example}).

Related perturbation problems in the context of opinion dynamics have been considered in \cite{acemoglu1} where the authors study a novel gossip consensus model where a limited number of pairwise interactions are asymmetric (one of the two agents engaged in the interaction, called forceful, does not change opinion). The mean behavior of agents is governed by a stochastic matrix $\tilde P$ which can be represented as the perturbation of a symmetric one $P$  (corresponding to the situation where all interactions are symmetric). Clearly, the consensus weight vector of $P$ is the uniform one $\pi_i=N^{-1}$ where $N$ is the number of nodes.
Their main results  (Theorems 5 and 6 therein) are explicit bounds of the distance between $\tilde \pi$ and $\pi$ in the infinity and in the $2$ norm. Connection with democracy can be obtained through the following straightforward inequalities
$$||\tilde\pi||_\infty-\frac{1}{N}\leq ||\tilde\pi-\pi||_{\infty}\leq ||\tilde\pi||_\infty+\frac{1}{N}$$
which implies that democracy can be equivalently expressed, in this context, by  $||\tilde\pi-\pi||_{\infty}\to 0$ (in correspondence of larger and larger graphs). Similar considerations also apply to the $2$-norm. 
Our results allow to conclude that when the set of forceful agents remains fixed and finite, the perturbed mean behavior remains democratic (see Theorem
\ref{theorempertdemtr} in this paper). This implies, in particular, that the perturbed invariant probability $\tilde\pi$ converges to the uniform one in the infinity norm. This convergence cannot be deduced in general 
directly from their Theorems 5 or 6 as their estimation contains a critical parameter at the denominator (the spectral gap in Theorem 5) which may be infinitesimal for certain families of graphs like grids. On the other hand, it is important also to remark that our result is only asymptotical and, differently from theirs, it does not lead to any explicit bound on the distance between consensus vectors.


See also \cite{acemoglu2} for related
results on the analysis of gossip consensus algorithms in the presence of
stubborn agents who never modify their opinion.

The type of perturbations discussed in this paper have also a
considerable importance in other contexts. When a consensus
algorithm is implemented into a real physical network \jcd{of communicating robots}, it is
possible that certain communications are down in one direction, or
that, in any case, messages are lost in one direction. Even if the
underlying stochastic matrix was designed to be reversible,
it is therefore possible that the actual algorithm will 
follow the dynamics of a perturbed stochastic matrix which is no longer
reversible. In this application it is important to avoid the situation where all agents converge to one single immobilized agent, resulting in a possible waste of energy. This case is easily ruled out by restricting our attention to perturbations that leave the network strongly connected, or the corresponding matrix irreducible, as it prevents any node from being stripped of all incoming edges. It is desirable to find supplementary conditions  ensuring that  if a  small number of communication channels break down, the final consensus position is not significantly far away from the arithmetic average of the agents' initial position, which is typically optimal in terms of resources.


Another application regards the webmaster problem \cite{Csaji,Tempo}: a webmaster has to choose the hyperlinks she puts on the
webpages she is responsible for in order to maximise their
PageRank, hence their visibility \jcd{on Web search engines}. \jcd{The
PageRank} is essentially the invariant distribution of a random
walker on the graph of hyperlinks \cite{BrinPage}, which is \jcd{described a stochastic matrix and therefore equivalent to a consensus problem}. 
While \cite{Csaji,Tempo} propose explicit algorithms to maximize the PageRank of a given page by choosing to rewire the hyperlinks leaving some webpages, we focus on an asymptotic situation where an ever-growing WorldWideWeb is called weakly democratic if it is impossible for a fixed small set of webpages to retain a fixed fraction of the total PageRank as more and more new pages are being added, and democratic if the top PageRank of the WWW keeps decreasing to zero as the network grows.

Let us briefly remark about the novelty of these results and their connections with classical Markov chains theory. As Proposition \ref{pos-rec} suggests, there is an intimate connection between weak-democracy of a sequence of stochastic matrices and the non-positive recurrence of the limit infinite stochastic matrix $P^{(\infty)}$ (precisely defined below). However, as Examples \ref{wk-not-dem} and \ref{demo-posrec} analyzed in Sections \ref{sec-weakdemo} and \ref{sec:demorecurrence} show, neither weak democracy nor democracy is  equivalent to the non-positive recurrence of $P^{(\infty)}$. While non-positive recurrence clearly plays an important role in our paper, there is nevertheless no obvious way to deduce our results from classical theory of infinite Markov chains. 

We remark that unlike the typical perturbation results
available in the literature where it is assumed that
$|P^{(n)}_{ij}-\tilde P^{(n)}_{ij}|$ are small, here we leave the
possibility of large perturbations but localized in a small set. For this type of perturbations, bounds like in \cite{Mitrophanov} (see Theorem 2.1) involving the reciprocal of the spectral gap of the matrix are of little utility since they will typically blow up when the number of nodes goes to $\infty$. A formula for updating the
invariant probability of a Markov chain upon the change of a row
of the probability transition matrix has been derived in
\cite{Meyer}, but there is no easy way to exploit it for an
asymptotic behavior.

\subsection{Outline of the paper}
 Section \ref{sec-formulation} is
devoted to introducing all relevant notation, to formulate the
problem,  to present some relevant examples, and to state the main results. In Section
\ref{sec-weakdemo} we study a weaker version of democracy when the
convergence to $0$ of the consensus weights (or invariant probabilities) is only
pointwise. We prove Theorem \ref{theoweakdemocracy} shows that under the
only assumptions that the limit chain is irreducible, weak
democracy is preserved under local perturbations. Moreover we show
with Example \ref{counter-example} that irreducibility of the
limit chain is a necessary condition for this type of results. In
Section \ref{sec:demorecurrence} we start analyzing democracy discussing its relation with non-positive recurrence. Our
main results are Theorem \ref{theoalllimits} which characterize
democracy in terms of the lack of positively recurrent states in
the asymptotic limits of the sequence, and Corollary
\ref{corpertdem} which establishes the preservation of democracy
under local perturbation. Finally, Section \ref{sec-democracy}
focuses on time-reversible chains and gives an application of
Corollary \ref{corpertdem}. The main result is Theorem
\ref{theorempertdemtr} which guarantees that democracy is preserved
under local perturbations even when these possibly break the
time-reversibility of the chain.

This paper extends a preliminary partial version that has appeared in conference proceedings \cite{Delvenne}. The main definitions and two main results of this paper, Theorems \ref{theoweakdemocracy} and  \ref{theorempertdemtr}, were already stated in \cite{Delvenne}. However in this paper we present complete and more elegant proofs, thanks in part to the new concepts of Section \ref{sec:demorecurrence}. \jcd{Example \ref{counter-example} and} Figure \ref{fig}, are reproduced from \cite{Delvenne}.


\section{Assumptions, examples, and main
results}\label{sec-formulation}

\subsection{Stochastic matrices and graphs} \label{sec:notations}
Given a set $V$ (finite or countably infinite), we denote by
$\1_V$ the vector in $\R^V$ having all components equal to $1$. A
stochastic matrix $P$ on $V$ is any $P\in \R^{V\times V}$ such
that $P_{ij}\geq 0$ for all $i,j\in V$ and $P\1_V=\1_V$.

To any stochastic matrix $P$ on a set $V$ (finite or infinite) we
can associate a directed transition graph ${\mathcal G}=(V,{\mathcal E})$
on the set of vertices $V$ and where $(i,j)\in {\mathcal E}$ if and
only if $P_{ij}>0$. If $i\in V$, $N(i)=\{\jcd{j} \in V\setminus\{i\}\,|\,
(i,j)\in V\}$ denotes the set of out-neighbors of $i$. Notice that
$i$ is never considered in $N(i)$ even when $(i,i)\in {\mathcal E}$.
A path on ${\mathcal G}$ is a sequence of vertices $\gamma =
(l_1,\dots , l_M)$ such that $(l_s,l_{s+1})\in{\mathcal E}$ for all
$s=1, \dots , M-1$. We say that $\gamma$ starts from $l_1$ and
ends in $l_M$, or, also, that joins $l_1$ to $l_M$. \jcd{The length of $\gamma$
 is denoted $l(\gamma):=M-1$}. \jcd{The graph} ${\mathcal G}$ is
said to be strongly connected if any pair of vertices can be
joined by a path; in this case the matrix $P$ generating the graph
is called irreducible. We will denote by $d_{\mathcal G}$ the usual distance on the vertices
of a strongly connected graph as the length of a minimal path
between vertices.
If $W\subseteq V$, we
put ${\mathcal G}(W)=(W,{\mathcal E}\cap (W\times W))$.

 If $P$ is an irreducible stochastic matrix
on a finite set $V$, it admits a unique vector $\pi\in\R^V$ with $\pi_i>0$ for all $i\in V$ such that
$\sum_i\pi_i=1$ and $\pi^*P=\pi^*$. If, moreover, $P$ is also aperiodic (see \cite{Levinetal} for the exact definition), it follows that $\lim_{n\to +\infty}P^n=\pi\1^*$: in this case, because of the interpretation presented in the Introduction, $P$ is also called a consensus matrix and the relative vector $\pi$ the consensus weight vector.

Any stochastic matrix \jcd{$P \in \R^{V\times V}$} can be interpreted as the transition matrix of a Markov chain on $V$. Given a probability vector $\rho\in\R^V$ ($\rho_i\geq 0$ for all $i\in V$ and $\sum_i\rho_i=1$), the pair $(\rho, P)$ is called a Markov chain and defines a stochastic process $X_t$ \jcd{(for times $t \in \N$) taking values in} $V$, \jcd{called the state space of the Markov chain. The initial state $X_0$ is distributed according to $\rho$ and the distribution of $X_{t+1}$ conditioned to $X_t=j$ is given by the $j$-th row of $P$.  This implies that $X_t$ on the state space $V$ is} given by $\rho^*P^t$. If $\rho$ is such that $\rho^*P=\rho^*$, it is said to be an invariant probability vector for $P$, and the corresponding Markov chain is said to be stationary. Consensus weight vectors can thus be interpreted as invariant probabilities. This dual Markov chain interpretation turns out to be a very powerful tool to state, prove and interpret our technical results and will thus be freely used in the following.

Given an undirected graph ${\mathcal G}=(V,{\mathcal E})$ (i.e., $(i,j)\in\mc E$ iff $(j,i)\in\mc E$) which is connected, an important class of stochastic matrices generating $\mc G$ can be constructed
starting from a symmetric non-negative valued matrix $C\in
\R^{V\times V}$ adapted to $\mc G$ (i.e.,  $C_{ij}\neq 0$ iff
$(i,j)\in\mc E$) and defining the stochastic matrix
\be\label{tr}P_{ij}=\frac{C_{ij}}{C_i}\ee where $C_i=\sum_jC_{ij}$
is assumed to be finite for all $i\in V$. A stochastic matrix of
this type is called time-reversible (or simply `reversible'), while $C$ is called a
conductance matrix. If $V$ is finite, the unique invariant
probability of $P$ is given by $\pi_{i}=C_i/\sum_j C_j$. In the
case when $C_{ij}\in\{0,1\}$ and $C_{ii}=0$ for all $i$, we obtain the matrix described in (\ref{SRW}) with $\tau=0$ which, in the probabilistic jargon, is called the simple random walk on $\mathcal G$.
Putting instead $C_{ii}=d_i\tau/(1-\tau)$ for all $i$, we obtain  the matrix described in (\ref{SRW}): for $\tau\neq 0$ this is called a lazy simple random walk.

\subsection{Assumptions and formulation of the
problem}\label{assumptions} We assume we have fixed an infinite
universe set ${\mathcal V}$ and a sequence $V_n$ \ff{($n\in\N$)} of finite
cardinality subsets of ${\mathcal V}$ \ff{which is nested (e.g. $V_n\subseteq V_{n+1}$) and is} such that $\cup_nV_n={\mathcal V}$. 
\ff{We then consider} a sequence of irreducible stochastic matrices $P^{(n)}$ on the
state spaces $V_n$ \ff{(and as such of increasing dimension)} with 
\ff{a property which essentially establishes that for every node $i\in{\mathcal V}$, the non-zero values of the $i$-th row of $P^{(n)}$ remain fixed for  $n$ sufficiently large. Formally}, we assume that, for
every $i\in\mc V$, there exists a positive integer  $n_i\in \N$  such that $i\in V_{n_i}$ and
\be\label{Pstabilize}P^{(n)}_{ij}=P^{(n_i)}_{ij}\,,\quad\forall
n\geq n_i\,,\;\forall j\in V_{n_i}\ee 
\ff{Notice first of all that in the formula above $i,j\in V_{n_i}\subseteq V_n$ for $n\geq n_i$ and thus \jcd{it} makes perfect sense to consider $P^{(n)}_{ij}$: the $i$-th rows  $P^{(n)}$ and $P^{(n_i)}$ are vectors of different lengths only differing \jcd{by zero entries}. \jcd{Therefore (\ref{Pstabilize}) means that the $i$th row of $P^{(n)}$ remains constant as soon as $n \geq n_i$, except for a growing string of zero entries.} If we consider the associated graphs $\mc G^{(n)}$ we have in particular that the out-neighbor\jcd{s} of node $i$ remain the same in all graphs for $n\geq n_i$.}
Property (\ref{Pstabilize}) allows us to
define, in a natural way, a limit stochastic matrix on ${\mathcal V}$.
For every $i,j\in \mc V$, we define
\be\label{infinityP}P^{(\infty)}_{ij}=\left\{\ba{ll}
P^{(n_i)}_{ij}\quad &{\rm if}\; j\in V_{n_i}\\ 0\quad &{\rm
otherwise}\ea\right.\ee Throughout the paper we will always use
the following notation convention. All quantities related to the
stochastic matrix $P^{(n)}$ (including $n=\infty$) will have the superscript ${(n)}$:
$\pi^{(n)}$ is the invariant probability \jcd{(uniquely defined for $n < \infty$ as we assume irreducibility of $P^{(n)}$)}, $\mc G^{(n)}$ the
associated graph, $N^{(n)}(i)$ the out-neighbor set of $i$ in $\mc
G^{(n)}$. Notice that, by (\ref{Pstabilize}), $\sum_{j\in V_{n_i}} P^{(n)}_{ij}=1$ for every $n\geq n_i$. This implies that $N^{(n)}(i)=N^{(n_i)}(i)$ for every $n\geq n_i$. In particular,
$N^{(\infty)}(i)$ is finite for all $i\in{\mathcal V}$.

\begin{definition}\label{maindef} The sequence of stochastic matrices $P^{(n)}$ is said to
be:
\begin{itemize}
\item weakly democratic if for all $i\in{\mathcal V}$, $\pi^{(n)}_i\to
0$ for $n\to +\infty$. \item democratic if
$||\pi^{(n)}||_{\infty}:=\max\limits_{i\in V_n}\pi^{(n)}_i\to 0$
for $n\to +\infty$.
\end{itemize}
\end{definition}

\medskip
In this paper we want to investigate the preservation of the
properties expressed in Definition \ref{maindef}, under finite
perturbations. More precisely, we fix a finite subset $W\subseteq
V_1$  and another sequence of irreducible stochastic matrices
$\tilde P^{(n)}$ on $V_n$ such that
\be\label{assumptionsPtilde}\ba{ll}\tilde
P^{(n)}_{ij}=P^{(n)}_{ij}\quad\forall i\in V_n\setminus W\,,\;\forall j\in V_n\\[8pt]
\tilde P^{(n)}_{ij}=\tilde P^{(1)}_{ij}\quad \forall\, i\in W\,,\;
\forall j\in V_1 \ea\ee In other terms, $\tilde P^{(n)}$ can be
seen as a perturbed version of $P^{(n)}$ with the perturbation
confined to the fixed subset $W$ and stable (it does not change as
$n$ increases). Notice that $\tilde P^{(n)}$ satisfy the same
stabilization assumption (\ref{Pstabilize}) than $P^{(n)}$, and
thus, also for this perturbed sequence we can define, following
(\ref{infinityP}), the asymptotic chain $\tilde P^{(\infty)}$. The
assumptions $W\subseteq V_1$ and the second one in
(\ref{assumptionsPtilde}) are taken for simplicity. The crucial
fact needed is that $W$ is finite and that for every $i \in W$ and
$j\in \mc V$,  $\tilde P^{(n)}_{ij}$ becomes constant for large
$n$.

\subsection{Examples}\label{sec-examples}
\ff{A general and fundamental example fitting in the formalism of \jcd{this section} can be obtained 
by starting from an infinite graph and considering simple random walks on larger and larger finite subgraphs of it. Precisely,
consider an infinite 
connected undirected graph $\mc G=(\mc V, \mc E)$ such that each
vertex $i\in \mc V$ has a finite degree $d_i$. Consider a nested sequence
$V_n$ of finite cardinality subsets of ${\mathcal V}$ such that $\cup_nV_n=\mc V$.  Assume that the subgraphs $\mc G^{(n)}=\mc
G(V_n)=(V_n, {\mathcal E}^{(n)})$ (where, we recall, $ {\mathcal E}^{(n)}=\mc E\cap V_n\times V_n$) are connected.  Notice that  $\mc G^{(\infty)}=\mc G$. Consider the lazy simple random walk \jcd{(\ref{SRW})} on it:  for $i,j\in V_n$ we put $P^{(n)}_{ij}=\frac{1-\tau}{d^{(n)}_i}$ for $(i,j)\in  \mc E^{(n)}$ with $i\neq j$ and $P^{(n)}_{ii}=\tau$. According to the notation agreement, $d^{(n)}_i$ denotes the degree of node $i$ in $\mc G^{(n)}$: clearly for sufficiently large $n$, this degree coincides with $d_i$.
Notice
that the invariant probability $\pi^{(n)}$ of $P^{(n)}$, is such that
\be\label{Pdemocratic}||\pi^{(n)}||_{\infty}\leq \frac{d^{(n)}
}{|{\mathcal E}^{(n)}|}\ee where $d^{(n)}:=\sup_{i\in
V_n}d_i$. In particular, this shows that if
$d^{(n)}=o(|{\mathcal E}^{(n)}|)$ for $n\to +\infty$, the sequence
$P^{(n)}$ is democratic.

The condition  $d^{(n)}=o(|{\mathcal E}^{(n)}|)$
is verified in all case\jcd{s} where the original graph has bounded
degrees (e.g. $d$-dimensional lattices and more general regular
graphs). Another case are the random geometric graphs: indeed at
the connectivity threshold, maximal degrees grow logarithmically
with the number of nodes, with high probability \cite{Penrose}.

\begin{example}\label{symmetric lattice} Consider the $d$-dimensional infinite lattice over ${\mathcal V}=\Z^d$ formally defined as follows. Consider the canonical basis vectors  $e_i\in\Z^d$ for $i=1,\dots ,d$ and put $\Lambda =\{\pm e_i\,|\,i=1,\dots ,d\}$. Then we define
$\mc G=(\Z^d, \mc E)$ where $\mc E:=\{(v,w)\in \Z^d\times\Z^d\;|\; v-w\in\Lambda\}$.
Consider $V_n=[-n,n]^d$. $\mc G^{(n)}=\mc
G(V_n)$ is the $d$-dimensional grid with $2n+1$ nodes in each direction. Internal nodes have degree $2d$ while boundary nodes have degrees in $\{1,2,\dots ,d\}$. In particular, $d^{(n)}=d$ for all $n$. This says that the corresponding simple random walks on such grid graphs form a democratic sequence.
\end{example}

A more general setting is obtainable by replacing the simple random walks with more general time-reversible matrices. Precisely, in the same graph setting proposed above, assume }
to have fixed a sequence of conductance matrices
$C^{(n)}$ adapted to $\mc G^{(n)}$ such that
\begin{enumerate} \item[(a)] for every $i\in\mc V$, there exist
$n_i\in \N$ such that $i\in V_{n_i}$ and
\be\label{Cstabilize}C^{(n)}_{ij}=C^{(n_i)}_{ij}\,,\quad\forall
n\geq n_i\,,\;\forall j\in V_{n_i}\ee \item[(b)] there exist
constants $0<a<b$ such that
\be\label{boundconductance}a<C^{(n)}_{ij}<b\,,\quad\forall
(i,j)\in {\mathcal E}^{(n)}\,,\;\forall n\in\N\ee
\end{enumerate}
Let $P^{(n)}$ be the time-reversible stochastic matrix on $V_n$
associated with $C^{(n)}$ in the sense of (\ref{tr}). Notice that
$P^{(n)}$ is irreducible and satisfies the stabilization condition
(\ref{Pstabilize}). Notice also that $P^{(\infty)}$ is
time-reversible and coincides with the stochastic matrix
associated with the limit of conductances $C^{(\infty)}$. 
\ff{A simple check on the invariant probabilities shows that the condition $d^{(n)}=o(|{\mathcal E}^{(n)}|)$ for $n\to +\infty$ remains sufficient for democracy.}

\bigskip
\ff{The following is instead a possible way to construct non
time-reversible examples, \jcd{as they occur on graphs that are not undirected}.

\begin{example}\label{exCayley} 

Consider the following modification of the grid graphs considered in Example \ref{symmetric lattice}. Similarly, we consider ${\mathcal V}=\Z^d$, the sequence of subsets $V_n=[-n,n]^d$, and the subset 
$\Lambda^+=\{e_i\,|\, i=1,\dots ,d\}$, \jcd{where $e_i$ is the elementary vector $(0,\ldots,0,1,0,\ldots,0)$ with a single $1$ entry in $i$th position.} We then define 
$\mc G^{(n)}=(V_n, \mc E^{(n)})$ where $\mc E^{(n)}:=\{(v,w)\in [-n, n]^d\times [-n, n]^d\;|\; v-w\in\Lambda^+\}$ where the $-$ operation is to be interpreted modulo $2n+1$. $\mc G^{(n)}$ is a $d$-dimensional grid with all directed edges without boundary (shaped like a \jcd{torus}).  \jcd{It is clearly non time-reversible as the walker may for instance jump from node $0$ to any node $e_i$ in one step, then to any node $e_i+e_j$, but not back to $0$.} It is easy to realize that $\mc G^{(\infty)}$ is the infinite lattice $\Z^d$ with all directed edges. \jcd{In the trivial case $d=1$, $\mc G^{(n)}$ is simply the $2n+1$-node directed cycle where the random walker runs always in the same direction, and $\mc G^{(\infty)}$ is the infinite directed path whose nodes are indexed by $\Z$.}
If we consider the simple random walk $P^{(n)}$
on $\mc G^{(n)}$, we have that the invariant probability is always
the uniform one, so that the sequence is democratic. Modifying the structure subset $\Lambda^+$, it is possible to construct a whole variety of graphs named Abelian Cayley graphs for which similar considerations apply.
\end{example}}



Both examples above lead to democratic sequence of stochastic matrices. We now
present examples of non weakly democratic sequence of stochastic matrices and
also weakly democratic sequences which are not democratic.

\begin{example}\label{wk-not-dem} Let $V_n=\{1,\dots ,n\}$, \jcd{$0 < \delta < 1$}, and
$$P^{(n)}_{ij}:=\left\{\ba{ll} \delta \quad &{\rm if}\; i<n\,,\;
j=i+1\\
1-\delta \quad &{\rm if}\; i>1\,,\;
j=i-1\\
1-\delta \quad &{\rm if}\; i=j=1\\
\delta \quad &{\rm if}\; i=j=n\ea\right.$$ It is \jcd{possible} 
to verify that, for $\delta\neq 1/2$, the invariant probability
measure is given by
$$\pi^{(n)}_i=\left(\frac{\delta}{1-\delta}\right)^{i-1}\frac{1-\left(\frac{\delta}{1-\delta}\right)}{1-
\left(\frac{\delta}{1-\delta}\right)^{n}}$$ If \jcd{$0 < \delta <1/2$}
it follows that

\begin{equation*}\lim\limits_{n\to +\infty}\pi^{(n)}_i=
\left(\frac{\delta}{1-\delta}\right)^{i-1}\left(1-\left(\frac{\delta}{1-\delta}\right)\right)\end{equation*}

so that the stochastic matrix is not
weakly democratic. Consider now the case when \jcd{$1/2 < \delta < 1$}.
Then,
$$\lim\limits_{n\to +\infty}\pi^{(n)}_i=0\quad \forall i$$
However,
$$\lim\limits_{n\to +\infty}||\pi^{(n)}||_{\infty}=\lim\limits_{n\to +\infty}\pi^{(n)}_n=1-\frac{1-\delta}{\delta}$$
so the stochastic matrix is weakly democratic but not democratic.
\end{example}

\subsection{Main results}

In this paper we will present two main results. The first one, the Theorem below, is a 
robustness result of weak democracy.

\begin{theorem}\label{theoweakdemocracy}
Consider a sequence of weakly democratic irreducible stochastic matrices
$P^{(n)}$ satisfying (\ref{Pstabilize}) with $P^{(\infty)}$
irreducible. Then, any perturbed sequence $\tilde{P}^{(n)}$ of
irreducible stochastic matrices satisfying (\ref{assumptionsPtilde}) and
such that $\tilde P^{(\infty)}$ is irreducible, is also weakly
democratic.
\end{theorem}

Theorem \ref{theoweakdemocracy} will be proven in Section \ref{sec-weakdemo}  where we will also present an example showing that irreducibility of the limit stochastic matrix is an assumption which can not be dropped.

The second result is within the framework of time-reversible stochastic matrices. We
have fixed an infinite universe set ${\mathcal V}$, a nested
sequence $V_n$ of finite cardinality subsets of ${\mathcal V}$ such that $\cup_nV_n=\mc V$, a
sequence of connected undirected graphs $\mc
G^{(n)}=(V_n, {\mathcal E}^{(n)})$, and a sequence of conductance
matrices $C^{(n)}$ adapted to $\mc G^{(n)}$ satisfying
(\ref{Cstabilize}). We impose two extra conditions. First, the
boundedness of the degrees on the infinite graph ${\mathcal
G}^{(\infty)}$: \beq\label{boundeddegree}d:=\sup_{i\in{\mathcal
V}}|N^{(\infty)}(i)|< +\infty\,.\eeq The second condition
strengthens (\ref{boundconductance}) and is a finiteness condition
on the values assumed by the conductance:
\be\label{assumptionsP}\Theta:=\{C^{(n)}_{ij}\,|\, i,j\in \mc
V\,,\; n\in\N\}\; \hbox{\rm is a finite set}\ee

Here is our main result (proof will be presented in Section \ref{sec-democracy}):

\begin{theorem}\label{theorempertdemtr} Consider a sequence of irreducible stochastic matrices
$P^{(n)}$ constructed through a sequence of conductance matrices
$C^{(n)}$ satisfying (\ref{Cstabilize}), (\ref{boundeddegree}),
and (\ref{assumptionsP}). Suppose that $P^{(\infty)}$ is
irreducible. Suppose moreover that the subset $W$ and the
perturbed sequence $\tilde P^{(n)}$ are chosen to satisfy
assumptions (\ref{assumptionsPtilde}) and $\tilde P^{(\infty)}$ is
irreducible. Then, the sequence $\tilde P^{(n)}$ is democratic.

\end{theorem}

The context of application of Theorem \ref{theorempertdemtr} is quite wide and includes many of the common cases which show up in consensus problems. \ff{We propose a couple of concrete instances below. The first one deals with  simple random walks on directed graphs obtained by finitely perturbing a sequence of undirected graphs (e.g cutting some edges in just one direction). The second instead models the presence of a finite community of nodes with homophilic behavior.}

\ff{\begin{example}\label{graph example}
Consider an infinite 
connected undirected graph $\mc G=(\mc V, \mc E)$ such that each
vertex $i\in \mc V$ has a finite degree. Assume that the subgraphs $\mc G^{(n)}=\mc
G(V_n)=(V_n, {\mathcal E}^{(n)})$ (where, we recall, $ {\mathcal E}^{(n)}=\mc E\cap V_n\times V_n$) are connected and consider the lazy simple random walk $P^{(n)}$ described in Section \ref{sec-examples}
$P^{(n)}_{ij}=\frac{1-\tau}{d^{(n)}_i}$ for $(i,j)\in  \mc E^{(n)}$ with $i\neq j$ and $P^{(n)}_{ii}=\tau$.
Clearly, (\ref{Cstabilize}), (\ref{boundeddegree}),
and (\ref{assumptionsP}) are all satisfied. Fix now any finite subset $W\subseteq V_1$ and consider a strongly connected perturbed graph (possibly no longer undirected) 
$\tilde{\mc G}=(\mc V, \tilde{\mc E})$ which can only differ 
from $\mc G$ for edges outgoing from $W$: for every $v\in \mc V\setminus W$ and $v'\in V$, it holds $(v,v')\in \mc E$ if and only if $(v,v')\in \tilde{\mc E}$. Assume that the subgraphs 
$\tilde{\mc G}^{(n)}=\tilde{\mc G}(V_n)$ are connected and consider the lazy simple random walk $\tilde P^{(n)}$ on $\tilde{\mc G}^{(n)}$ formally defined as $\tilde P^{(n)}_{ij}=\frac{1-\tau}{\tilde d^{(n)}_i}$ for $(i,j)\in  \tilde{\mc E}^{(n)}$ with $i\neq j$ and $\tilde P^{(n)}_{ii}=\tau$, where $\tilde d^{(n)}_i$ is the number of outgoing neighbors of $i$ in the graph $\tilde G^{(n)}$.
Theorem \ref{theorempertdemtr} can be applied to conclude that $\tilde P^{(n)}$ is a democratic sequence.
\end{example}}

\ff{\begin{example}\label{main example}
Let $\mc G$, $\mc G^{(n)}$ and $P^{(n)}$ as in the previous example. Given a  finite subset $W\subseteq V_1$, define the perturbed sequence $\tilde{P}^{(n)}$ as follows

\begin{equation}\label{pertSRW}\tilde P^{(n)}_{ij}=\left\{\begin{array}{ll}\frac{\lambda(1-\tau)}{d^{(n)}_i+(\lambda -1)d^{(n)}_{i,W}}\quad &\hbox{\rm  \jcd{for $i\in W$, $j\in W$ neighbor of $i$}}\\
\frac{1-\tau}{d^{(n)}_i+(\lambda -1)d^{(n)}_{i,W}}\quad &\hbox{\rm  \jcd{for $i\in W$, $j\notin W$ neighbor of $i$}}\\
\tau\quad &\hbox{\rm  for $j=i$}
\end{array}\right.
\end{equation}
where $d^{(n)}_{i,W}$ is  the number of neighbors of $i$ inside $W$, and $\lambda \jcd{\geq} 1$ measures the homophily  of community $W$, \jcd{measuring a strong mutual influence inside $W$ and weak influence by the agents outside $W$ in an opinion dynamics interpretation}. 

\begin{figure}  
\begin{center}
\includegraphics[width=8cm, trim=3.5cm 9.5cm 4cm 10cm, clip]{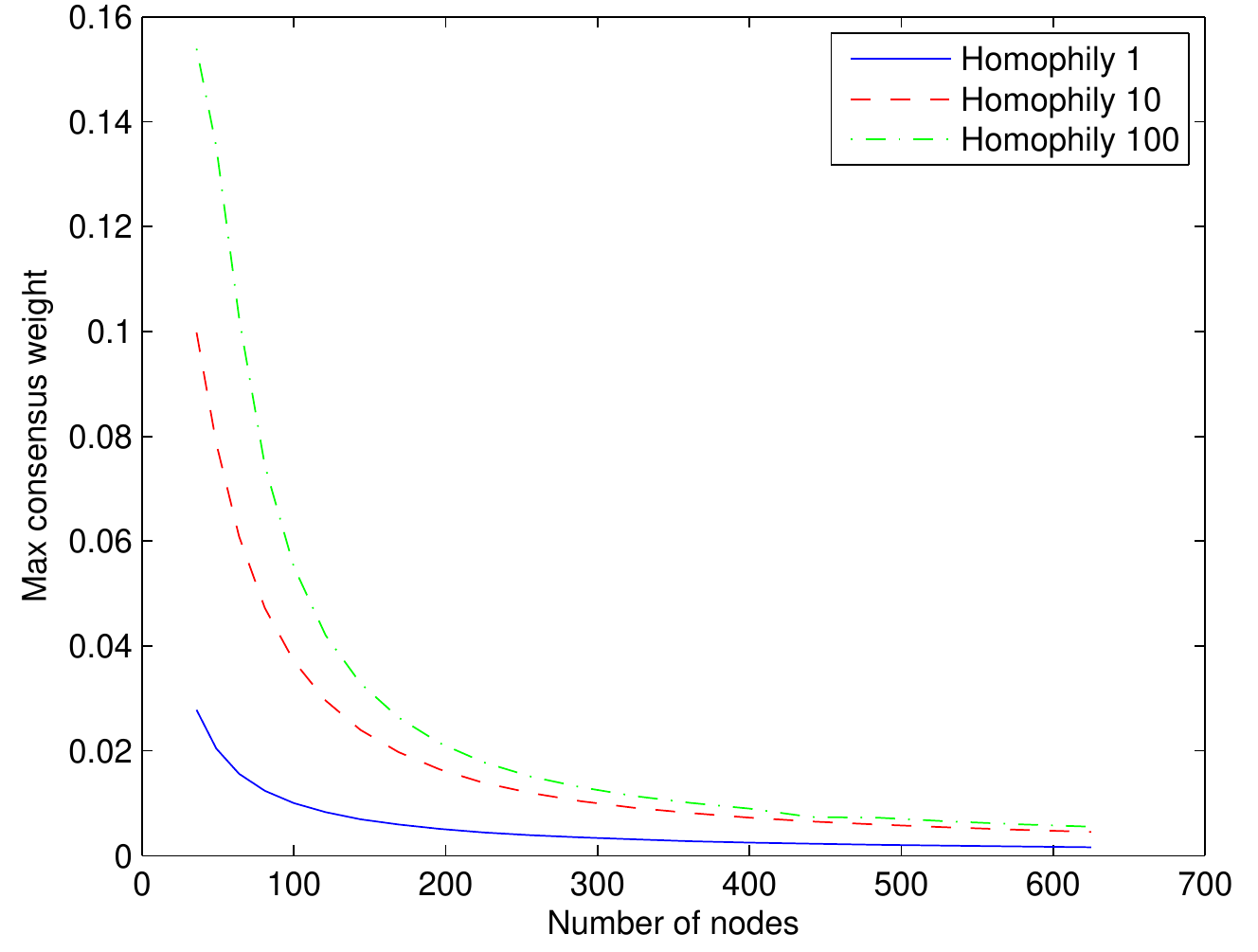} \hfill\\
\end{center}   \caption{\jcd{We illustrate our main result, Theorem   \ref{theorempertdemtr}, on Example \ref{graph example} applied to a bidimensional torus grid, whose node set is of the form $[-n,n]^2$, as in Example \ref{exCayley}. We consider a lazy random walk with self-confidence parameter $\tau=0.1$. We then modify the behavior of the 9-node community $W=[-1,1]^2$, with a homophily parameter $\lambda=1$, $10$ or $100$. We plot the maximum consensus weight $\|\pi^{(n)}\|_{\infty}$ as a function of the number of nodes. Note that homophily $\lambda=1$ corresponds to the unperturbed lazy random walk, where all nodes have equal weight. A high homophily creates a community $W$ of nodes of much higher consensus weight than the other nodes. As predicted by Theorem \ref{theorempertdemtr}, this homophilic community, despite breaking time-reversibility of the random walk, does not break democracy. As the number of nodes increases indeed, the maximum weight converges to zero.}}\label{fighomo}
\end{figure}

Theorem \ref{theorempertdemtr} says that the minority in $W$, even if their homophily parameter $\lambda$ is very large, cannot unilaterally break the democracy: their consensus weights $\tilde\pi_i$ will go to $0$ in the large scale limit and therefore also their opinion, \jcd{regardless of} their conservativeness, will have a  negligeable impact in the asymptotic consensus opinion of the global population. See illustration on Figure \ref{fighomo}.
\end{example}}

\section{Weakly democratic stochastic matrices and hitting
times}\label{sec-weakdemo} 

\ff{
In this section we prove Theorem \ref{theoweakdemocracy}. Techniques employed, as it will happen in the following sections, are essentially probabilistic. 
The key point is the connection between weak democracy and hitting and return times of the underlying Markov chain.

For the sake of completeness, we briefly recall below a number of standard concepts of Markov chains which will play an important role in the following analysis.
\jcd{Recall from Section \ref{sec-formulation} that a Markov chain on a state space $V$ is a stochastic process $(X_t)$ described by an initial probability vector $\rho\in\R^V$ describing the distribution of $X_0$ and a stochastic matrix $P\in\R^{V\times V}$ describing the transition probabilities from $X_t$ to $X_{t+1}$.}
We will denote by $\P_i$ the probability relative to such a process when $\rho=\delta_i$ the
delta probability measure concentrated on $i$. Similarly we denote
by $\E_i$ the corresponding mean operator. If $S\subseteq
V$, $\tau_S$ and $\tau_S^+$ denote, respectively, the first
hitting time and the first return time into $S$:
$$\begin{array}{rcl}\tau_S&:=&\min\{t\geq 0\,|\, X_t\in S\}\\ \tau_S^+&:=&\min\{t\geq
1\,|\, X_t\in S\}\;\; (X_0\in S)\,.\end{array}$$  If
$S=\{i\}$ we use the notation $\tau_i$ and $\tau_i^+$ for $\tau_S$
and $\tau_S^+$, respectively. For notation simplicity we will use
the notation $E_{ij}$ for $\E_i(\tau_j)$ and
$E_{i+}$ for $\E_i(\tau^+_i)$. 
In the case when $P$ is irreducible and $\pi\in\R^V$ is its corresponding invariant probability, we have the following remarkable relation:
$\pi_i=(E_{i+})^{-1}$ for all $i\in V$ (see for instance \cite{AldousFill,Levinetal}).

We now assume we have fixed a sequence of stochastic matrices $P^{(n)}$ and a
perturbed one $\tilde P^{(n)}$ as in Section \ref{assumptions}. According to our general terminological assumptions we will denote the above quantities with a superscript $(n)$ when referred to 
$P^{(n)}$ and with in addition a tilde on top when referred to the perturbed one (e.g. $E^{(n)}_{ij}$, $E^{(n)}_{i+}$, $\tilde E^{(n)}_{ij}$, $\tilde E^{(n)}_{i+}$).

Since $\pi^{(n)}_i=(E^{(n)}_{i+})^{-1}$, it follows that a sequence of stochastic matrices $P^{(n)}$
is democratic if and only if $\lim_{n \to \infty} E^{(n)}_{i+}= \infty$ for all $i$. 
}

The following theorem is proved in \cite{Delvenne}.

\begin{proposition} \cite{Delvenne} \label{propwd}
For a sequence of irreducible stochastic matrices $P^{(n)}$ satisfying
(\ref{Pstabilize}) and such that $P^{(\infty)}$ is irreducible,
the following conditions are equivalent:

\begin{enumerate}
\item[(a)] The sequence is weakly democratic. \item[(b)] There
exists $j\in {\mathcal V}$ such that $$E^{(n)}_{j+}\to +\infty\;\;{\rm
for}\; n\to +\infty$$ \item[(c)] There exist $j\in{\mathcal V}$ and a
finite subset $Y\subseteq {\mathcal V}\setminus\{j\}$ such that
$$\max\limits_{i\in Y}E^{(n)}_{ij} \to \infty\;\;{\rm for}\; n \to
\infty$$ \item[(d)] For every $j\in{\mathcal V}$ there exists a finite
subset $Y\subseteq {\mathcal V}\setminus\{j\}$ such that
$$\max\limits_{i\in Y}E^{(n)}_{ij} \to \infty\;\;{\rm for}\; n \to
\infty$$
%
%
\end{enumerate}
\end{proposition}

\medskip
\noindent {\bf Remark:} It follows from Proposition \ref{propwd}
that when weak democracy fails, then $\pi^{(n)}_i$ remains bounded
away from $0$ for all $i$. This phenomenon is essentially due to
the stabilizing condition (\ref{Pstabilize}) which prevents nodes'
degrees to blow to $\infty$. See \cite{Golub} for examples where
instead this condition is not imposed.

\medskip

We are now ready to prove Theorem \ref{theoweakdemocracy}.

\proof (of Theorem \ref{theoweakdemocracy})  We first prove it in the case when $W=\{w\}$, i.e., when only one node $w$ is perturbed. Applying Proposition
\ref{propwd} (condition (d)) to $P^{(n)}$ we have that there
exists $Y\subseteq {\mathcal V}\setminus\{w\}$ such that
$$\max\limits_{i\in Y}E^{(n)}_{iw} \to \infty\;{\rm for}\; n \to
\infty$$ Notice that $E^{(n)}_{iw}=\tilde{E}^{(n)}_{iw}$ for every
$i\in Y$. Hence, condition (c) of Proposition \ref{propwd} is
verified for $\tilde P^{(n)}$. Hence $\tilde P^{(n)}$ is weakly
democratic.

Consider now a general, finite set of perturbed nodes $W$. The idea is to use induction on the cardinality of $W$. However, some attention must be paid to the possible loss of irreducibility of $P^{(\infty)}$ during the inductive path. To overcome this, we consider three cases.

\begin{enumerate}

\item Assume that for every node $w \in W$, the set
of outgoing edges of $w$ in the graph of $P^{(\infty)}$ is included in the set of outgoing edges of $w$ in the graph of $\tilde{P}^{(\infty)}$.
For every $W' \subseteq W$, construct the sequence of stochastic matrices $P^{(n)}_{W'}$, obtained from $P^{(n)}$ by replacing every row corresponding to $w \in W'$ by the corresponding row of $\tilde{P}^{n}$. In particular, $P^{(n)}_{W}=\tilde{P}^{(n)}$. Then, obviously every $P^{(\infty)}_{W'}$ is irreducible. A straightforward inductive procedure applied to a sequence of $W'$ where one node is added at a time now allows to prove that $\tilde{P}^{(n)}$ is weakly democratic.

\item Now assume that for every node $w \in W$, the set
of outgoing edges of $w$ in the graph of $P^{(\infty)}$ contains the set of outgoing edges of $w$ in the graph of $\tilde{P}^{(\infty)}$.
Then every $P^{(\infty)}_{W'}$ is irreducible again and, arguing as in previous case, we conclude that $\tilde{P}^{(n)}$ is weakly democratic.

\item If none of the above apply, consider the intermediate sequence of chains $Q^{(n)}= \frac{1}{2} (P^{(n)} + \tilde{P}^{(n)})$. Then the first case above applies to $P^{(n)}$ and $Q^{(n)}$, showing that $Q^{n}$ is weakly democratic. Now the second case applied to $Q^{(n)}$ and $\tilde{P}^{(n)}$ shows that $\tilde{P}^{(n)}$ is weakly democratic.\jcd{\qed}
\end{enumerate}

We give now an example of a weakly democratic family of stochastic matrices $P^{(n)}$, converging to an irreducible infinite stochastic matrix,
such that, modifying the transition probabilities from just one
state, one gets a non weakly democratic family. This is not in
contradiction with Theorem \ref{theoweakdemocracy}, because this
perturbed family converges to a reducible infinite stochastic matrix.
This shows that the irreducibility assumption on $\tilde{P}^{\infty}$ is required.

\begin{example}\label{counter-example}

The chains $P^{(n)}$ and $\tilde{P}^{(n)}$ are defined as in
Figure \ref{fig} (notice that $P^{(n)}$ fits in Example
\ref{exCayley}). It is clear that the stationary distribution on
states of $P^{(n)}$ is uniform, therefore $P^{(n)}$ is weakly
democratic. We shall now show that the sequence of chains
$\tilde{P}^{(n)}$ is not weakly democratic.

Indeed $\tilde{E}^{(n)}_{0+}=
1+\tilde{E}^{(n)}_{10}=1+E^{(n)}_{10}$. The last equality follows
from the fact that computing the first hitting time from 1 to 0
does not require the knowledge of the transition probabilities
from 0. Moreover, $E^{(n)}_{10}=E^{(\infty)}_{1,\{0, n-1\}}$, by
construction of $P^{(n)}$. Note that evaluating
$E^{(\infty)}_{1,\{0, n-1\}}$ is equivalent to the classical
gambler's ruin problem (random walk with a drift and two absorbing
barriers) \cite{Feller}. It follows that $E^{(\infty)}_{1,\{0,
n-1\}} <c$, where $c$  is a finite constant, independent of $n$.
Therefore,  $\tilde{\pi}^{(n)}_0 > 1/c$. Thus, the sequence
$\tilde{P}^{(n)}$ is not weakly democratic. With the same technique, one may show that the probability $\tilde{\pi}_k^{(n)}$ for vertices $k>0$ does not converge to zero, while $\tilde{\pi}_k^{(n)} \to 0$ for any $k<0$.

\begin{figure} 
\begin{center}
\includegraphics[width=6cm]{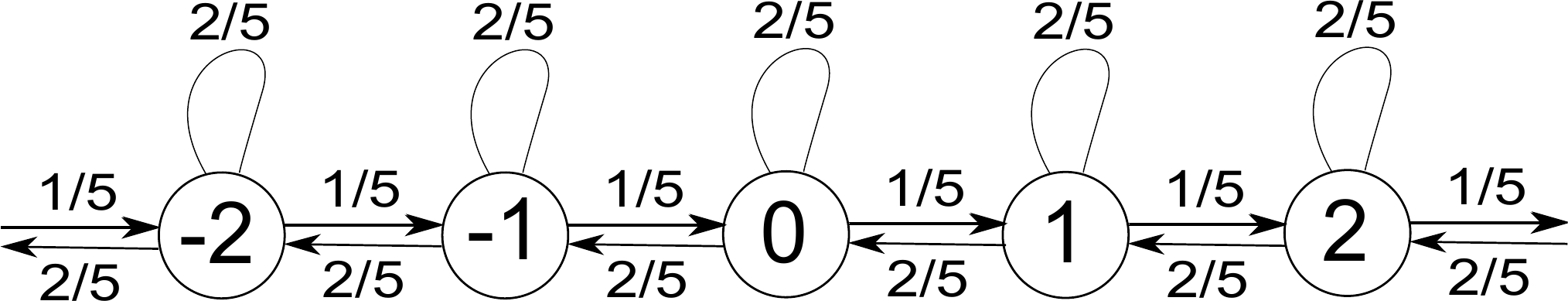} \hfill\\
\hfill \\
\includegraphics[width=6cm]{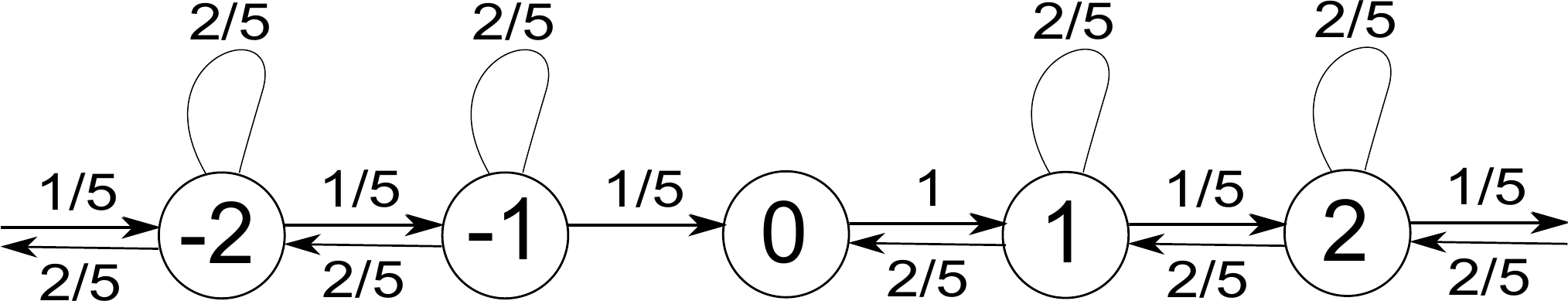}
\end{center}   \caption{The chains of Example \ref{counter-example}. \emph{Top}:
the chain $P^{(\infty)}$. The chain $P^{(n)}$ is built from
$P^{(\infty)}$ by identifying $\lfloor n/2 \rfloor $ and $- \lceil
n/2 \rceil$, thus making $P^{(n)}$ a cyclic chain. \emph{Bottom}:
the chain $\tilde{P}^{(\infty)}$, which differs from
$P^{(\infty)}$ only by transitions from state 0. The chains
$\tilde{P}^{(n)}$ differ from $P^{(n)}$ in the same way.}\label{fig}
\end{figure}
\end{example}

\section{Weak democracy, democracy and
recurrence}\label{sec:demorecurrence}

There are important connections between weak democracy and democracy
of a sequence $P^{(n)}$ and the positive recurrence of the limit
stochastic matrix $P^{(\infty)}$. We recall that a vertex $i$ is said
to be positively recurrent if $E^{(\infty)}_{i+}<+\infty$
\cite{Woess}. If the matrix is irreducible and one vertex is
positively recurrent, then all vertices are positively recurrent
and we say, in this case, that the chain is positively recurrent.
We have the following simple relation:

\begin{proposition}\label{pos-rec} Consider a sequence of irreducible stochastic matrices
$P^{(n)}$ such that both $P^{(n)}$ and its transpose $(P^{(n)})^*$ satisfy (\ref{Pstabilize}). Assume moreover that $P^{(\infty)}$
does not have positively recurrent vertices. Then, the sequence
$P^{(n)}$ is weakly democratic.
\end{proposition}   
\proof The proof follows by standard arguments. Indeed, a diagonal
argument shows that we can always find a subsequence $n_k$ such
that $\pi^{(n_k)}_i$ converges to some value that we denote $\pi^{(\infty)}_i$ for all $i\in {\mathcal
V}$ and a dominated convergence argument shows that
$\pi^{(\infty)\,*}P^{(\infty)}=\pi^{(\infty)\,*}$ (where the ${.}^*$ denotes the matrix transposition). Now, if
$P^{(n)}$ is not weakly democratic, $\pi^{(\infty)}_i$ is
different from $0$ for all $i$. Hence, $P^{(\infty)}$ admits an
invariant probability measure and thus
$E^{(\infty)}_{i+}<+\infty$. \qed

Let us now go back to Example \ref{wk-not-dem} to show that non
positive recurrence does not yield democracy. Indeed
$P^{(\infty)}$ in that case is given by
$$P^{(\infty)}_{ij}:=\left\{\ba{ll} \delta \quad &{\rm if}\;
j=i+1\\
1-\delta \quad &{\rm if}\; i>1\,,\;
j=i-1\\
1-\delta \quad &{\rm if}\; i=j=1\ea\right.$$  It is straightforward
to verify that, for \jcd{$0< \delta <1/2$}, $P^{(\infty)}$ admits an
invariant probability measure  given by
$$\pi^{(\ff{\infty})}_i=\left(\frac{\delta}{1-\delta}\right)^{i-1}\left(1-\left(\frac{\delta}{1-\delta}\right)\right)$$
Therefore the chain is positively recurrent. Instead, if
\jcd{$1/2 < \delta < 1$}, $P^{(\infty)}$ does not admit any invariant
probability measure, the chain is thus non positively recurrent.
Nevertheless, the sequence $P^{(n)}$ is only weakly democratic and
not democratic.

We now present another example showing that $P^{(n)}$ may well be
democratic while $P^{(\infty)}$ is positively recurrent. The
reason is that there can be `boundary effects' coded in the
sequence $P^{(n)}$ which disappear in the limit\jcd{, in the sense that the consensus weight vector of $P^{(n)}$ is largely determined by the many entries in $P^{(n)}$ that have not yet stabilized to their eventual value in $P^{(\infty)}$, as we illustrate in the next example}.

\begin{example}\label{demo-posrec}
Fix an infinite universe set ${\mathcal V}$, a nested sequence
$V_n=\{1,\dots ,n\}$ of finite cardinality subsets of ${\mathcal V}\jcd{=\N}$,  
a sequence of strongly connected graphs $\mc G^{(n)}=(V_n, {\mathcal
E}^{(n)})$ equipped with simple random walks $P^{(n)}$ for which
the usual stability condition applies. Independently from
$P^{(n)}$ being democratic or not, we now show that we can modify
it to make it democratic without changing the limit matrix
$P^{(\infty)}$. Consider the
modified sequence of graphs $\tilde{\mc G}^{(n)}=(V_n\cup A_n,
\tilde{\mathcal E}^{(n)})$ where \jcd{$A_n=\{n+1,\dots ,n+M_n\}$} and
$\tilde{\mathcal E}^{(n)}=\jcd{{\mathcal E}^{(n)}\cup\{(n, n+1), (n+1,
n+2), \dots ,(n+M_n, 1)\}}$. \ff{The sequence $M_n$ will be chosen later.}
Let $\tilde P^{(n)}$ be the
simple random walk on $\tilde{\mc G}^{(n)}$. For every $i\in V_n$,
let $\gamma_i^{(n)}$ be any simple path in $\mc G^{(n)}$ connecting
$i$ to \jcd{$n$} and let $q^{(n)}_i$ be \jcd{the product of probability of the various edges composing the path}. Starting from a vertex $i\in V_n$, with probability $q^{(n)}_i$ we will be following path $\gamma_i^{(n)}$ to \jcd{$n$} and with probability \jcd{$1/(d_{n}+1)$} we will then choose the edge connecting to \jcd{$n+1$}. At that point we will be forced to follow the \jcd{whole} directed cycle of length $M_n$ up to \jcd{$1$}. Hence,
\begin{equation}\label{bound1}\tilde\E_{i}^{(n)}(\tau^+_{i})\geq \frac{q^{(n)}_iM_n}{\jcd{d_{n}+1}}\end{equation}
Similar considerations show that, if $i\in A_n$, for sure we have
\begin{equation}\label{bound2}\tilde\E_{i}^{(n)}(\tau^+_{i})\geq M_n\end{equation}
\ff{If we choose $M_n$ to be 
$$M_n:=\frac{\jcd{d_{n}}+1}{(\min\limits_{i\in V_n}
q^{(n)}_i)}n$$
we obtain from (\ref{bound1}) and (\ref{bound2}) that}
$$\min_{i\in V_n\cup
A_n}\tilde\E_{i}^{(n)}(\tau^+_{i})=+\infty$$
 so that the sequence of modified chains is indeed democratic. Notice that
$P^{(\infty)}= \tilde P^{(\infty)}$ (nodes in $A_n$ disappear in
the limit). If we have started from a positively recurrent
$P^{(\infty)}$ (one can consider for instance the sequence in
Example \ref{wk-not-dem} with \jcd{$0 < \delta  <1/2$}) we have now
constructed an example of a democratic sequence whose limit chain
is positively recurrent.
\end{example}

\medskip

The negative results presented above show that the limit chain by
itself is not sufficient to capture the property of democracy of a
chain. The reason, as already noticed, are the `border' effects
which disappear in the limit. However, as will be shown below, the
non positive recurrence will play a role once we consider
different limit notions for the sequence of chains, to include
border effects.

We now introduce a concept which will play a central role in the
following. We start fixing some notation. If $\mc G=(V,\mc E)$ is
a graph, $i\in V$ and $R>0$ denote $B_{\mathcal G}(i, R)=\{j \in
V\;|\; d_{\mathcal G}(i,j)\leq R\}$. For simplicity we will use the
notation ${\mathcal G}(i,R)={\mathcal G}(B_{\mathcal G}(i, R))$.

Given a sequence of irreducible stochastic matrices $P^{(n)}$ satisfying
(\ref{Pstabilize}), we say that a stochastic matrix $Q$ on an infinite
set ${\mathcal Z}$ is a {\it limit } of $P^{(n)}$ if there exist\jcd{s}
a subsequence of positive integers $n_k$, a sequence $l_{k}\in
V_{n_k}$, and a sequence of injective maps $\lambda_k:V_{n_k}\to
{\mathcal Z}$, such that
\begin{enumerate} \item[(a)] $\lambda_k(l_{k})=z_0$ constant.
\item[(b)] For every $z\in\mc Z$, there exists $k_0\in\N$
such that $z\in \lambda_k(V_{n_k})$ for all $k\geq k_0$.
\item[(c)] For every integer $R>0$ there exists $k_0$ such that:
for every $k\geq k_0$, for every $i,\,j\in B_{{\mathcal
G}^{(n_k)}}(l_{k},R)$, we have that
\beq\label{limitpoint}P^{(n_k)}_{ij}=Q_{\lambda_k(i)\lambda_k(j)}\eeq
\end{enumerate}
\ff{Notice that properties above essentially assert that the sequence of graphs ${\mathcal
G}^{(n_k)}$ become\jcd{s} stable in an arbitrary fixed neighborhood subgraph of $l_k$ and isomorphic to $\mc Z$ in a neighborhood of $z_0$. Moreover, $P^{(n_k)}$ and $Q$ are equal for large $k$ in such neighborhood subgraphs.}

\ff{Clearly, $P^{(\infty)}$ is always a limit of $P^{(n)}$: it
is sufficient to choose any constant sequence $l_n$. Such a limit
point is called {\it trivial}. 

The sequence $l_{k}$ appearing in the definition above is called a {\it stabilizing} sequence \jcd{and is said to be trivial if constant}.}

\jcd{For example, consider $P^{(n)}$ describing the simple random walk on the square grid $V_n=[-n,n]^2$, as in Example \ref{symmetric lattice}. The trivial limit is the simple random walk on $\Z^2$. However, choosing $l_n$ as the node of coordinate $(0,-n)$ leads to a limit stochastic matrix $Q$ that describes the random walk on $\Z \times \N$, while $l_n=(-n,n)$ leads to a simple random walk on $\N \times \N$.}

$P^{(n)}$ will be called {\it complete} if for any sequence
$l_n\in V_n$, there always exists a stabilizing subsequence.

We have the following result which generalizes Proposition
\ref{pos-rec}.

\begin{theorem}\label{theoalllimits} Consider a sequence of irreducible stochastic matrices
$P^{(n)}$ satisfying (\ref{Pstabilize}) which is weakly
democratic, complete, and such that every non-trivial limit 
does not contain positively recurrent points. Then, the sequence
$P^{(n)}$ is democratic.
\end{theorem}
\proof
Let $\pi^{(n)}$ be the invariant probability of $P^{(n)}$. Suppose
by contradiction that
$$\lim\limits_{n\to +\infty}\sup_{i\in V_n}\pi^{(n)}_i>0$$
Then, there exists a subsequence of the positive integers $n_k$
and a sequence $i_k\in V_{n_k}$ such that \beq\label{contr}
\pi^{(n_k)}_{i_k}\geq \alpha>0\,\;\;\forall k\in \N\eeq 
\ff{If $i_k$ admits a constant subsequence (e.g. $i_k=l$ for infinite values of $k$, then (\ref{contr}) would violate weak democracy.
Therefore, by completeness,  we can assume with no lack of generality that $i_k$ is stabilizing and non trivial.}
This
guarantees the existence of a sequence of embeddings
$\lambda_k:V_{n_k}\to {\mathcal Z}$ \ff{and a stochastic matrix $Q$ on $\mc Z$ satisfying properties (a), (b), and (c) \jcd{just above}. Let $\mc H$ be the graph induced by the matrix $Q$.}
Fix $R>0$ and consider the events
$$L_k(R)=\{\tau^+_{i_k}<\tau_{B_{{\mathcal G}^{(n_k)}}(i_k, R)^c}\}$$
$$\Lambda(R)=\{\tau^+_{z_0}<\tau_{B_{{\mathcal H}}(z_0, R)^c}\}$$
\ff{(where the notation $A^c$ denotes the complementary of subset $A$).
Choose now $k$ large enough so that  property (b)
holds true for every $z\in B_{{\mathcal H}}(z_0, R)$ and also property (c) holds. 
This implies that for such $k$, the two stochastic processes governed by $P^{(n_k)}$ and by $Q$, respectively,  and starting from $i_k$ and $z_0$, respectively, have the same statistics (through the embedding $\lambda_k$) as long they are in the balls $B_{{\mathcal G}^{(n_k)}}(i_k, R)$ and $B_{{\mathcal H}}(z_0, R)$, respectively.
This yields}
\beq\label{dicot1}\begin{array}{rcl}\E^{(n_k)}_{i_k}(\tau^+_{i_k})
&\geq&\E^{(n_k)}_{i_k}(\tau^+_{i_k}\1_{L_k(R)})+R\P^{(n_k)}_{i_k}(L_k(R)^c)\\
&=&\E^{Q}_{z_0}(\tau^+_{z_0}\1_{\Lambda(R)})+R\P^{Q}_{z_0}(\Lambda(R)^c)\end{array}
\eeq where $\P^{Q}_{z_0}$ and $\E^{Q}_{z_0}$ denote probability
and mean with respect to the chain $Q$ starting from $z_0$. From
(\ref{contr}) and (\ref{dicot1}) \ff{(recalling that $\pi^{(n_k)}_{i_k}=\E^{(n_k)}_{i_k}(\tau^+_{i_k})^{-1}$)} we then obtain \beq\label{dicot2}
\E^{Q}_{z_0}(\tau^+_{z_0}\1_{\Lambda(R)})+R\P^{Q}_{z_0}(\Lambda(R)^c)\leq
\alpha^{-1}\,,\;\forall R>0\eeq Observing that, for $R\to
+\infty$, $\Lambda(R)\uparrow\{\tau^+_{z_0}<+\infty\}$, we 
obtain that
\begin{eqnarray} &\label{limitsP}\lim\limits_{R\to +\infty}\P^{Q}_{z_0}(\Lambda(R)^c)=\P^{Q}_{z_0}(\tau^+_{z_0}=+\infty) \\ 
&\label{limitsE}\lim\limits_{R\to
+\infty}\E^{Q}_{z_0}(\tau^+_{z_0}\1_{\Lambda(R)})=
\E^{Q}_{z_0}(\tau^+_{z_0}\1_{\tau^+_{z_0}<+\infty})\end{eqnarray}
\ff{From (\ref{limitsP}) and (\ref{dicot2}) we obtain that $\P^{Q}_{z_0}(\tau^+_{z_0}=+\infty)=0$. Considering (\ref{limitsE})  and  using again (\ref{dicot2}) we finally obtain $\E^{Q}_{z_0}(\tau^+_{z_0})<+\infty$, namely that $z_0$ is positively recurrent. This contradicts the standing assumptions of the theorem. The result is thus proven.} \jcd{\qed}

A slight variation of the argument used to prove Theorem
\ref{theoalllimits} allows to prove  a perturbation result.

\begin{corollary}\label{corpertdem} Consider a sequence of weakly democratic irreducible stochastic matrices
$P^{(n)}$ satisfying (\ref{Pstabilize}) and such that  $P^{(\infty)}$ is irreducible, the sequence is complete, and every
non-trivial limit chain does not contain positively recurrent
points.  Suppose moreover that the subset $W$ and the perturbed
sequence $\tilde P^{(n)}$ are chosen to satisfy assumptions
(\ref{assumptionsPtilde}) and $\tilde P^{(\infty)}$ is also
irreducible. Then, the sequence $\tilde P^{(n)}$ is democratic.
\end{corollary}
\proof
Notice first of all that because of Theorem
\ref{theoweakdemocracy}, $\tilde P^{(n)}$ is weakly democratic.
Notice now that all non-trivial limit chains of $\tilde P^{(n)}$
coincide with the non-trivial limits chains of $P^{(n)}$. Hence
result follows from Theorem \ref{theoalllimits}. \jcd{\qed}

\section{Perturbation of time-reversible democratic
chains}\label{sec-democracy}

In this section we present some applications of Corollary
\ref{corpertdem}. In particular, we will prove Theorem \ref{theorempertdemtr}.




\medskip

We start with the following result:

\begin{proposition}\label{boundary}
Suppose (\ref{Cstabilize}), (\ref{boundeddegree}), and
(\ref{assumptionsP}) hold true. Then, the corresponding sequence
$P^{(n)}$ is complete and all its limit chains are non positively
recurrent chains.
\end{proposition}

\proof
Recall the notation ${\mathcal G}(i,R)$  to denote the subgraph of
$\mc G$ consisting of those vertices whose distance from $i$ is
not greater than $R$. If $C$ is a conductance matrix on $\mc G$,
we denote by $C(i,R)$ its restriction to $B_{\mathcal G}(i, R)$.

Consider, preliminarily,  the set $\Omega_{d,R, \Theta}$ of all
triples $({\mathcal G}, C, x)$ where ${\mathcal G}=(V,{\mathcal E})$ is an
undirected graph with degrees bounded from above by $d$, $C$ is a
conductance matrix adapted to $\mc G$ and taking values in
$\Theta$, and $x\in V$ is such that $d_{\mathcal G}(x,v)\leq R$ for
every $v\in V$. In particular, the cardinality of the vertices of
such graphs is bounded from above by $d^R$. On
$\Omega_{d,R,\Theta}$ we can introduce a notion of isomorphism:
$({\mathcal G}, C, x)$ and $({\mathcal H}, C^\prime, y)$ are called
isomorphic (denoted $({\mathcal G}, C, x)\sim({\mathcal H}, C^{\prime},
y)$) if there exists a graph isomorphism $\psi:{\mathcal G}\to {\mathcal
H}$ such that $C_{ij}=C^{\prime}_{\psi(i), \psi(j)}$ for all $i$
and $j$, and, moreover, $\psi(x)=y$. It is evident (recall that
$\Theta$ is a finite set) that $\sim$ is an equivalence relation
and that the set of equivalence classes $\Omega_{d,R,
\Theta}/\sim$ is finite.

Notice now that for any fixed sequence $l_n\in V_n$, and for every positive number $R$,
$({\mathcal G}^{(n)}(l_n,R),  C^{(n)}, l_n) \in \Omega_{d, R,
\Theta}$. Hence, there exists a subsequence $l_{n^R_k}$  such that
$({\mathcal G}^{(n^R_k)}(l_{n^R_k},R), C^{(n^R_k)}(l_{n^R_k},R),
l_{n^R_k})$ all belong to the same equivalence class. Denote by
$({\mathcal H}^{(R)}, D^{(R)}, z^{(R)})$ a fixed representative in
$\Omega_{d, R, \Theta}$ of such class. A straightforward inductive
argument shows that the subsequences $n_k^R$ can be chosen in such
a way that, if $R_1<R_2$, then $n_k^{R_2}$ is a subsequence of
$n_k^{R_1}$. If this is the case, then, necessarily, if $R_1<R_2$,
we have that $({\mathcal H}^{(R_2)}(z^{(R_2)}, R_1),
D^{(R_2)}(z^{(R_2)}, R_1), z^{(R_2)})$ is isomorphic to $({\mathcal
H}^{(R_1)}, D^{(R_1)}, z^{(R_1)})$. Considering the direct limit
with respect to $R$ of such triples $({\mathcal H}^{(R)}, D^{(R)},
z^{(R)})$, we thus obtain the existence of a graph ${\mathcal H}$ with
a conductance matrix $D$ adapted to it and with a vertex $z_0$
such that, for every $R>0$, $({\mathcal H}(z_0, R), D(z_0, R), z_0)$
is isomorphic to $({\mathcal H}^{(R)}, D^{(R)}, z^{(R)})$. It is a
standard fact that $D$ induces a not positively recurrent
stochastic matrix $Q$. As a final step, it is sufficient to choose
$n_k=n_k^k$ to complete the proof. \jcd{\qed}

This allows to prove our main result:

\proof (of Theorem \ref{theorempertdemtr})
Proposition \ref{boundary} insures that Corollary \ref{corpertdem}
can be applied. The result then follows. \jcd{\qed}




We conclude discussing an example where we do not have time-reversibility.
\begin{example}[Example \ref{counter-example} revisited]
The sequence $P^{(n)}$ in Example \ref{counter-example} is complete and any limit chain is simply the biased random walk on the bi-infinite line which is known to be transient (see \cite{Levinetal}). The reason why $\tilde P^{(n)}$ in Example \ref{counter-example} is actually non democratic follows from the fact that $\tilde P^{(\infty)}$ is not irreducible. Any perturbation involving a finite subset $W$ which keeps $\tilde P^{(\infty)}$ irreducible, will thus be democratic because of Corollary \ref{corpertdem}.
\end{example}

\section{Conclusions and further problems}

In this paper, we have discussed the concept of democracy for
sequences of opinion dynamics and, using the language of stochastic matrices, we have given results guaranteeing the preservation of democracy under finite local perturbations.
There are many issues which remain open and which, in our opinion,
deserve future attention. The following is a partial list of them:
\begin{enumerate}
\item[(a)] It would be of interest to estimate the rate of
convergence to $0$ of the infinity norm of the invariant
probability of the perturbed sequence considered in Theorem
\ref{theoalllimits}. The proof proposed does not allow for a
straightforward estimation and new ideas are probably needed;

\item[(b)] What happens if the set $W$ is grows unboundedly but remains `small enough' with
respect to $V_n$? 

\item[(c)] Stronger variants democracy can be explored. For example we can request bounded ratios
$\pi^{(n)}_i/\pi^{(n)}_j$, implying that $\pi^{(n)}_i\asymp
1/n$ for $n\to +\infty$. It would be interesting to be able to
generalize our results to this stronger notion of democracy.

\end{enumerate}

\section{Acknowledgement}

The authors thank Federica Garin and Julien Hendrickx for
their contributions on the Example \ref{counter-example} and
Giacomo Como and Sandro Zampieri for useful discussions on the
problems studied in this paper. This article presents research
results of the Belgian Programme on Interuniversity Attraction
Poles, initiated by the Belgian Federal Science Policy Office and the ARC `Large graphs and Networks' of the French Community of Belgium. The
scientific responsibility rests with the authors.


\thebibliography{1}
\bibitem{acemoglu1} D. Acemoglu, A. Ozdaglar, A. Parandeh Gheibib  ``Spread of (mis)information in social networks'', \emph{Games and Economic Behavior}
Vol.~70(2), pp. 194-227,~2010.

\bibitem{acemoglu2} D. Acemoglu, G. Como, F. Fagnani, A. Ozdaglar,
Opinion fluctuations and disagreement in social networks, available on-line at http://arxiv.org/abs/1009.265


\bibitem{AldousFill} D. Aldous, J. Fill, ``Reversible Markov chains and random walk on graphs'',
book in preparation, draft available on-line at
http://www.stat.berkeley.edu/~aldous/RWG/book.html

\bibitem{BrinPage}
S. Brin and L. Page, ``The anatomy of a large-scale hypertextual {W}eb search
         engine'', \emph{Computer Networks and ISDN Systems}, vol. 30(1-7), pp. 107-117, 1998.


\bibitem{Carli} R. Carli, F. Fagnani, A. Speranzon, S. Zampieri,
``Communication constraints in the average consensus problem'', \emph{Automatica} vol.~44(3),
pp.~671-684, 2008.

\bibitem{Csaji}
B. C. Cs\'aji and R. M. Jungers and V. D. Blondel, ``PageRank optimization by edge selection'',
submitted, 2009

\bibitem{Delvenne} F. Fagnani, J.-C. Delvenne,  ``Democracy in Markov chains and its preservation under local perturbations'', \emph{Proceedings 49th IEEE-CDC Conference, Atlanta}, pp. 6620 - 6625 , ~2010.


\bibitem{Feller}
William Feller, {\em An introduction to probability theory and its
applications} (2nd edition), vol.\ I, Wiley, 1971

\bibitem{Frasca}
P. Frasca, C. Ravazzi, R. Tempo, and H. Ishii, ``Gossips and Prejudices: Ergodic Randomized Dynamics in Social Networks'',
\emph{arXiv preprint}, arXiv:1304.2268, 2013.


\bibitem{Friedkin}
N.E. Friedkin and E.C. Johnsen, ``Social influence networks and opinion change'',
\emph{Advances in Group Processes}, vol. 16(1), pp. 1-29, 1999.




\bibitem{Golub} B. Golub, M. O. Jackson, `` Na\"{\i}ve Learning in Social Networks and the Wisdom of Crowds ''
\emph{American Economic Journal: Microeconomics}, vol. 2(1), pp.
112-49, 2010.

\bibitem{Tempo}
H. Ishii and R. Tempo, ``Distributed Randomized Algorithms for the PageRank Computation'',
\emph{IEEE Transactions on Automatic Control}, vol. 55, pp. 1987-2002, 2010.


\bibitem{Jackson} M. O. Jackson, {\em Social and economic
networks}, Princeton University Press, 2008.

\bibitem{Jadba} A. Jadbabaie, J. Lin, and A. S. Morse, ``Coordination of groups
of mobile autonomous agents using nearest neighbor rules''
\emph{IEEE Transactions on Automatic Control}, vol. 48(6), pp.
988-1001, 2003.

\bibitem{Levinetal}
David A. Levin, Yuval Peres and Elizabeth L. Wilmer, {\em Markov
Chains and Mixing Times}, American Mathematical Society, 2008.
Available online at
http://www.uoregon.edu/~dlevin/MARKOV/markovmixing.pdf


\bibitem{Meyer}
Carl D. Meyer and James M. Shoaf. ``Updating finite Markov chains by using techniques of
group matrix inversion''. {\em Journal of Statistical Computation and Simulation}, vol. 11, pp. 163-181, 1980.

\bibitem{Mitrophanov}
A. Yu Mitrophanov ``Stability and exponential convergence of continuous-time Markov chains''. {\em Journal of Applied Probability}, vol. 40, pp. 970-979, 2003.

\bibitem{Olfati} R. Olfati-Saber, J. A. Fax, and R. M. Murray, ``Consensus and
cooperation in networked multi-agent systems'', \emph{Proceedings
of the IEEE}, vol. 95(1), pp. 215-233, 2007.

\bibitem{Penrose} M. Penrose, {\em Random Geometric Graphs}, Oxford Studies in
Probability. Oxford Univ. Press, 2003.

\bibitem{Woess} W. Woess, {\em Random walks on infinite graphs and
groups}, Cambridge University Press, 2000.
\end{document}